\documentstyle[11pt,newpasp,twoside]{article}
\def\edcomment#1{\iffalse\marginpar{\raggedright\sl#1\/}\else\relax\fi}
\reversemarginpar

\begin{document}
\title{A test of arm induced star formation in spiral galaxies from near-IR and H$\alpha$ imaging}
\author{M.S. Seigar$^1$ \& P.A. James$^2$}
\affil{{\footnotesize{\em $^1$Astrophysics Group, Department of Physics, Imperial College, Prince Consort Road, London SW7 2BZ, U.K.\\
$^2$Astrophysics Research Institute, Liverpool John Moores University, Egerton Wharf, Birkenhead CH41 1LD, U.K.}}}

\begin{abstract}
We have imaged a sample of 20 face-on spiral galaxies in H$\alpha$
and in the near-IR K-band (2.2$\mu$m), in order to
determine the location and strength of star formation
with respect to perturbations in the old stellar population. We
found that star formation rates are significantly enhanced
near K-band arms. If K-band light is
dominated by emission from old stars, this shows that
density waves trigger star formation.
However, no significant correlation was found between
the global underlying disk properties of spiral galaxies
 and their total star formation rates.  
\end{abstract}

\vspace*{-0.5cm}
\section{Introduction}

There are two models that describe star formation in spiral galaxies. The 
first of these in the Large Scale Shock Scenario (e.g. Roberts 1969) in 
which stars are formed when gas is compressed by shocks caused by density 
waves in spiral arms. The response of the gas to the shock can be non-linear 
if the relative motion is supersonic. The alternative is Stochastic Self 
Propagating Star Formation (e.g. Seiden \& Gerola 1982) in which stars form 
randomly, e.g. when gas clouds compressed by shocks from supernovae. 
Density waves are still used in this model to concentrate 
the star formation in 
spiral arms. In this paper we test these models using a sample of 20 
galaxies and methods described below.

\section{Observations and Data Reduction}

20 spiral galaxies were observed in the K-band with IRCAM3 on UKIRT and in 
H$\alpha$ with the CCD camera on the JKT. Detected H$\alpha$ fluxes were 
converted to intrinsic luminosities using a distance calculated from the 
heliocentric recession velocity and $H_0=75$kms$^{-1}$Mpc$^{-1}$. 
Star formation rates were derived from the H$\alpha$ luminosities using 
the empirical correlation of Kennicutt (1998).

\section{Results}

We tested to see whether the properties of the spiral arms, as
measured in the near-IR (Seigar \& James 1998), have any impact on
the total rate of disk star formation, as traced by H$\alpha$
emission. We find no significant correlations 
between disk H$\alpha$ luminosity
and arm strength, parametrized by a seeing-independent `Equivalent Angle' (EA)
(Seigar \& James 1998). Dividing EA by the FWHM of arm
cross-section gives a measure of arm contrast, which plausibly
determines the strength of shocks induced in the inter-stellar gas by
the spiral arm. No correlation was found with with disk H$\alpha$ luminosity. 
We then found that H$\alpha$ 
surface brightness anticorrelates with arm EA and arm 
contrast. Thus, strong spiral arms do
not lead to more intense star formation in these galaxies. This is in
agreement with Elmegreen \& Elmegreen (1986), who found
no difference in overall star formation rates in galaxies with and
without strong spiral structure.

However, arms may have a local effect
on star formation rate within the disks of individual galaxies,
although other factors may dominate differences in star formation 
rate between galaxies. 
We thus came up with a test, based
on a technique used to measure star-formation efficiency (Knapen et al.
1996), to identify arm-induced star formation in these
galaxies. For galaxies with well-defined spiral arms
in their K-band images, we rebinned the K images to the same
pixel scale as the H$\alpha$ images, and overlaid them, 
using stellar centroids as references.  
A polygon lying around each K-band arm was defined. Regions
were also fitted by eye to define the disk, excluding
the central bulge. The counts
within each of the arm regions, and the overall disk, were
measured.  These apertures were then imported into the
H$\alpha$ images and photometry was then obtained for the same regions. 
Finally, the H$\alpha$/K flux ratio was calculated
for all of the areas. The test is
to see whether this ratio is larger in the arms than for
the disks, where the latter ratio is taken from the total
disk minus the central region. If arms represent regions where 
all disk material is concentrated by
an equal factor, then the H$\alpha$/K ratio would be the same in arms
as in the disk.  However, in most cases, we
found H$\alpha$/K ratios to be significantly higher in the K-band arms
than in the disks. 37 arm regions were measured in
14 galaxies; 27 of these arms had H$\alpha$/K
ratios greater than that of the disk of the same galaxy. We find 
that 13 of the 14 galaxies show a net enhancement in
H$\alpha$/K ratio in spiral arms. The mean ratio found for all galaxies
shows that H$\alpha$/K is 1.4$\pm$0.1 times higher in the arms
than in the corresponding disks.  This 
represents a highly significant finding of triggered 
star formation within spiral arms. 

\section{Conclusions}

We have found an increase in H$\alpha$ flux near K-band 
arms in this sample. If K-band arms are dominated by light from 
old stars, this can be interpreted as star formation triggered by a 
density wave. Globally, no significant correlation was found 
between the properties of the old stellar disk and star formation
rates.


\begin{references}
\reference Elmegreen B.G., Elmegreen D.M., 1986, ApJ, 311, 554
\reference Kennicutt R.C., 1998, ApJ, 498, 541
\reference Knapen J.H., Beckman, J.E., Cepa J., Nakai N., 1996, A\&A, 308, 27
\reference Roberts W.W., 1969, ApJ, 174, 859
\reference Seiden P.E., Gerola H., 1982, Fundam. Cosmic Phys., 7, 241
\reference Seigar M.S., James P.A., 1998, MNRAS, 299, 685
\end{references}
\end{document}